\documentclass[aps,showpacs]{revtex4}
\usepackage{graphicx}

\newcommand{\ket}{\rangle}
\begin{document}
\title{ Quantum Game with Restricted Matrix Strategies}
\author{Bo Chen$^{1,2}$, Ying-Jun Ma$^{1,2}$  and Gui-Lu Long$^{1,2,3,4}$
\thanks{Corresponding author:gllong@tsinghua.edu.cn}}
\address{ $^1$ Department of Physics, Tsinghua University, Beijing 100084, China \\
$^2$ Key Laboratory For Quantum Information and Measurements, Beijing 100084, China\\
$^3$ Center for Atomic and Molecular NanoSciences, Tsinghua University, Beijing 100084,China\\
$^4$ Institute of Theoretical Physics, Chinese Academy of
Sciences, Beijing 100080, China}
\date{\today}

\begin{abstract}
We study a quantum game played by two players with restricted
multiple strategies.  It is found that in this restricted quantum
game Nash equilibrium does not always exist when the initial
state is entangled.  At the same time, we find that when Nash
equilibrium exists the pay off function is usually different from
that in the classical counterpart except in some special cases.
This presents an explicit
 example where quantum game and classical game may differ. When
 designing a quantum game with limited strategies, the allowed strategy should be
 carefully chosen according to the type of initial state.
 \end{abstract}
 \maketitle
 Quantum game theory has become a new area of application of
 quantum theory. Two-player quantum
 game\cite{r1,r2}, multi-player non-cooperative quantum game \cite{r3} and the cooperative
 three player quantum game \cite{r4,r5} have been reported recently. In these studies,
 only  $2\times 2$ matrix strategy have been considered\cite{r1,r2,r3,r4,r5,r6,r7,r8}
 where the two players have only two pure strategies respectively and the payoff
 matrix of player is a two-dimensional matrix. In classical game theory\cite{r9},
 $2\times 2$ matrix strategy game is only a small part in the whole game theory. It is
 therefore natural to generalize game theory in quantum mechanics with multiple
 strategies.
 In this work, we will consider
 the $N\times M$ $(N,M\ge 2)$ matrix strategy where one player has $N$ and the other
 has $M$ pure strategies.

 The rules of the quantum game are as follows: 1) before the game starts, the
 initial state are known to the players, just like the case in a classical game.
  In our work, we suppose the initial state has the form
   of $a_{0}|0\ket+\ldots+a_n|n\ket$ with $a_0$, $a_{1}\ldots a_{n}\ge 0$.
   This initial state can be a product state or
   an entangled state; 2) each player makes a unitary transformation on the initial state
   to place his strategy. By varying the parameter in the unitary transformation, he can choose
   his strategy for his benefit.

  We begin by simple examples and then generalize the results into higher dimensions.


 Suppose player has two pure strategies and player B has three pure strategies.
 The payoff matrix of player $A$ is given by
     \begin{equation}
     \left[ \begin{array}{rrrr} & \left[ 1\right] & \left[ 2\right] & \left[3\right] \\
     \left[ 1\right] & 2 & 3 & -2 \\ \left[ 2\right] & -2 & 4 & 2 \end{array} \right]
     \label{1} \end{equation}
 where, the columns index $[i]$ denotes the pure strategies
      of player B , and the row  index $[j]$ is the strategies for player A.
      In classical game,
      $Nash$ $Equilibrium$ can be  obtained by mixed strategies, where player A chooses his
      two strategies with equal probability: $(\frac{1}{2},\frac{1}{2})$.
      His payoff is zero. Similarly the mixed strategy for player B
       is $(\frac{1}{2},0,\frac{1}{2})$,  and payoff is zero too.   In quantum
       game, the players take their strategies by changing the quantum state of the game machine
       using a unitary operation.  A general unitary transformation can be written as
         \begin{equation} U=\left( \begin{array}{cc} e^{-i\frac{\alpha }{2}} & 0 \\ 0 &
         e^{i\frac{\alpha }{2}} \end{array} \right) \left( \begin{array}{cc}
         \cos \frac{\beta }{2} & -\sin \frac{\beta }{2} \\ \sin \frac{\beta }{2} &
         \cos \frac{\beta }{2} \end{array} \right). \label{e2} \end{equation}
         We restrict ourselves into the following unitary transformation
         $U=\sqrt{p}I+i\sqrt{1-p}\sigma _{x}$, where by changing the parameter $p$, we can take different
         strategies.

         For player $A$, he can choose in principle any operation in the $U(3)$ group. But we
         can set some restrictions which corresponds to different rules of the game. For instance,
         in three-dimensional space,
         the following   unitary operation
         \begin{equation} U_{3}=R_{z}(\alpha )R_{y}
         (\beta )R_{z}(\gamma )=\left( \begin{array}{ccc} \cos \alpha & -\sin \alpha  &
          0 \\ \sin \alpha  & \cos \alpha  & 0 \\ 0 & 0 & 1 \end{array} \right) \left(
          \begin{array}{ccc} \cos \beta  & 0 & \sin \beta  \\ 0 & 1 & 0 \\ -\sin \beta
          & 0 & \cos \beta \end{array} \right) \left( \begin{array}{ccc} \cos \gamma  &
           -\sin \gamma  & 0 \\ \sin \gamma & \cos \gamma  & 0 \\ 0 & 0 & 1 \end{array}
           \right),   \label{e3} \end{equation}
           will be one possible choice. However, this restricted unitary operator is still
            very complicated because it contains 3 parameters.  To simplify our discussion, we require
            the unitary operation to depends on only one parameter. Furthermore, the operator
            can make superposition of all the pure strategies which is possible in quantum game,
            but is not possible in classical game. The purpose is to see the effects that brings about
            by a quantum game machine. We find that the following operator
\begin{eqnarray}
U_{3}=\sqrt{q} I-\sqrt{1-q}
 \exp (i\theta )M,
 \end{eqnarray}
 where
 \begin{eqnarray}
 \theta &=& \arccos (\frac{1}{2}\sqrt{(1-q)/2q},\\
  M& =&\frac{1}{\sqrt{2}}\left[ \begin{array}{ccc} 0 & 1 & 1 \\ 1 & 0 & 1 \\
 1 & 1 & 0 \end{array} \right],
  \label{e4}
  \end{eqnarray}
 is unitary and can produce a superposition of the various strategies.
By choosing a different $q$, Bob chooses different strategy. Now, let us choose the
initial
 state as $|\psi _{in}\rangle =a_{0}|11\rangle +a_{1}|22\rangle $,\ $(\left\vert a_{0}\right
 \vert ^{2}+ \left\vert a_{1}\right\vert ^{2}=1)$. Without loss of
 generality, we suppose $ a_{0},a_{1}\ge 0$. By applying
 unitary matrices to the initial state,  the output state of the game machine becomes
 \begin{equation}
 |\psi _{out}\rangle =c_{1}|11\rangle +c_{2}|12\rangle +c_{3}|13\rangle
 +c_{4}|21\rangle + c_{5}|22\rangle +c_{6}|23\rangle \label{5}
 \end{equation}
 where
 \begin{eqnarray}
 \left\vert c_{1}\right\vert ^{2} &=&a_{0}^{2}pq+\frac{1}{2}
 a_{1}^{2}(1-p)(1-q)  +\sqrt{2}\sin (\theta )a_{0}a_{1}\sqrt{pq(1-p)(1-q)}, \label{e6} \\
  \left\vert c_{2}\right\vert ^{2} &=&\frac{1}{2} a_{0}^{2}p(1-q)+  a_{1}^{2}q(1-p)
  -\sqrt{2}\sin (\theta )a_{0}a_{1}\sqrt{ pq(1-p)(1-q)},  \nonumber\\
  \left\vert c_{3}
  \right\vert ^{2} &=&\frac{1}{2}a_{0}^{2}p(1-q)+  \frac{1}{2} a_{1}^{2}(1-p)(1-q),
  \nonumber \\
  \left\vert c_{4}\right\vert ^{2} &=&a_{0}^{2}q(1-p)+  \frac{1}{2} a_{1}^{2}
  p(1-q)-\sqrt{2}\sin (\theta )a_{0}a_{1}\sqrt{pq(1-p)(1-q)}, \nonumber \\
  \left\vert c_{5}
  \right\vert ^{2} &=&\frac{1}{2} a_{0}^{2}(1-p)(1-q)+a_{1}^{2}pq+  \sqrt{2}\sin (\theta )
  a_{0}a_{1}\sqrt{ pq(1-p)(1-q)}, \nonumber \\  \left\vert c_{6}\right\vert ^{2} &=&
  \frac{1}{2}a_{0}^{2}(1-p)(1-q)+  \frac{1}{2 }a_{1}^{2}p(1-q).  \nonumber
  \end{eqnarray}
  We can obtain the payoff function of player A as follows
  \begin{equation}
   P_{A}=(\frac{13}{2}a_{0}^{2}+a_{1}^{2})pq+5\sqrt{2}
  \sin (\theta )a_{0}a_{1} \sqrt{pq(1-p)(1-q)}-\frac{5}{2}a_{0}^{2}p+(3a_{1}^{2}-5a_{0}^{2})q
  +3a_{0}^{2}. \label{e7}
  \end{equation}
  When $a_{0}=1$, $a_{1}=0$, where no entanglement is
   present,  the payoff function of   player A is
   \begin{equation}
   P_{A}=\frac{13}{2}pq-
   \frac{5}{2}p-5q+3.  \label{e8}
   \end{equation}
   The payoff of player A with respect to $p$ and
   $q$ is plotted in Fig.1. The 3D-figure displays a shape of saddle, and the saddle-point
   is the \emph{Nash Equilibrium} point. The precise value can be found by solving
   \begin{eqnarray}
   \frac{\partial P_{A}}{\partial p} &=&\frac{13}{2}q-\frac{5}{2}=0 \label{e9}
    \\
    \frac{\partial P_{A}}{\partial q} &=&\frac{13}{2}p-5=0, \nonumber
    \end{eqnarray}
    which give the \emph{Nash Equilibrium} strategies at $p=\frac{10}{13}$, $q=\frac{5}{13}$.
     The corresponding payoff is $\frac{14}{13}$, and payoff for player $B$
     is $-\frac{14}{13}$. If $a_{0}=0$, $a_{1}=1$ , then the payoff of player A is $0$
     in \emph{Nash
     Equilibrium} strategies, which is exactly the same as that in
     classical game. In these two cases, the initial state is a product states, a Nash equilibrium
     can be found. But it is not always equal to the classical counterpart.

     It is interesting to point out that although in classical game Nash equilibrium always
     exist, it is not true in an arbitrarily designed quantum game such as this one.
     When entanglement is present in
     the initial state, the quantum game may have quite different properties compared with
     classical game. If we choose an entangled initial state, say
     let $a_{0}=a_{1}=\frac{1}{\sqrt{2}}$, no \emph{Nash Equilibrium}
     point exists. This example clearly shows the difference between classical game and
     quantum game. When the initial state is a product, Nash equilibrium can be found, but the
     payoff may not be the same as that in the classical game.

Suppose both players have three distinct strategies,   and the payoff matrix of
player A is
 \begin{equation}
 \left[ \begin{array}{cccc} & \left[ 1\right] &
 \left[ 2\right] & \left[ 3\right]\\ \left[ 1\right] & 2 & 0 & 2 \\ \left[ 2\right] & 0
  & 3 & 1 \\ \left[ 3\right] & 1 & 2 & 1 \end{array} \right]. \label{e10}
  \end{equation}
  The classical \emph{Nash Equilibrium} strategy is  ($\frac{1}{3},0,
  \frac{2}{3}$) for player A, and $(\frac{1}{3},\frac{1}{3},\frac{1}{3})$ for player B.
  At \emph{Nash Equilibrium}, the payoff of player A is $\frac{4}{3}$, and
   player B $-\frac{4}{3}$. In quantum game theory, we let the initial state
   $|\psi _{in}\rangle =a_{0}|11\rangle +a_{1}|22\rangle +a_{2}|33\rangle $ where
   $(\left\vert a_{0}\right\vert ^{2}+\left\vert a_{1}\right\vert ^{2}+\left\vert a_{2}
   \right \vert ^{2}=1;a_{0},a_{1},a_{2}\ge 0)$. The players take the operators
   $U_{3A}= \sqrt{p}I-\sqrt{1-p}\exp (i\theta _{1})M$,
   $\theta _{1}=\arccos (\frac{1}{2}\sqrt{(1-p)/2p})$ and $U_{3B}
   =\sqrt{q}I-\sqrt{1-q}\exp (i\theta _{2})M$, where $ \theta _{2}=\arccos (\frac{1}{2}
   \sqrt{(1-q)/2q})$, respectively.
   The payoff function of  player $A$ is
    \begin{eqnarray}
   P_{A} &=&\frac{5}{4}+(3a_{0}a_{2}+2a_{1}a_{2}+3a_{1}a_{0})\cos (\theta _{1}+ \theta _{2})
   \sqrt{pq(1-p)(1-q)}  \label{11} \\ &&+(2a_{0}a_{2}+3a_{1}a_{2}+3a_{1}a_{0})\cos
   (\theta _{1}-\theta _{2})\sqrt{ pq(1-p)(1-q)} \nonumber\\ &&-(2a_{1}a_{0}+a_{0}a_{2}+
   2a_{1}a_{2})(1-p)\cos (\theta _{2})\sqrt{q(1-q)} \nonumber \\ &&-(a_{1}a_{2}+2a_{0}a_{2}
   +3a_{1}a_{0})(1-q)\cos (\theta _{1})\sqrt{p(1-p)}+ \frac{1}{4}a_{1}^{2}+
   \frac{1}{2}a_{0}^{2}  \nonumber \\ &&+(a_{1}a_{2}+\frac{1}{2}a_{1}a_{0}+\frac{3}{2}
   a_{0}a_{2})(1-pq)+(3a_{1}^{2}+ \frac{9}{4}a_{0}^{2}-\frac{3}{4}a_{2}^{2})pq \nonumber
   \\ &&+(\frac{1}{4}a_{2}^{2}-\frac{3}{2}a_{0}a_{2}-a_{1}a_{2}-a_{1}^{2}-\frac{3}{ 4}
   a_{0}^{2})(p+q)+ \frac{1}{2}(a_{1}^{2}-a_{0}^{2})q. \nonumber
   \end{eqnarray}
   If the initial state has no entanglement, say $a_{0}=1$, $a_{1}=a_{2}=0$,
   then the payoff function of  player A is
   \begin{equation}
   P_{A}=\frac{9}{4}pq-\frac{3}{4}p-\frac{5}{4}q+\frac{7}{4}. \label{e12}
   \end{equation}
   We have plotted the 3d-figure in Fig.2.
   We get a saddle point at $p=\frac{5}{9 }$, $q=\frac{1}{3}$. The payoff function
   $P_{A}=\frac{4}{3}$ in \emph{Nash Equilibrium} strategy which is the same as that
   of classical game. However, when $a_{0}=a_{1}=a_{2}=\frac{1}{\sqrt{3}}$,
   the payoff function is
   \begin{eqnarray}
   P_{A} &=&\frac{1}{2}pq-\frac{4}{3}p-\frac{4}{3}q+\frac{5}{2}+\frac{16}{3}
   \cos \theta _{1}\cos \theta _{2}\sqrt{pq(1-p)(1-q)}  \label{e13} \\
   &&-\frac{5}{3}(1-p)\cos \theta _{2}\sqrt{q(1-q)}-2(1-q)\cos \theta _{1}\sqrt{
   p(1-p)}.
   \nonumber
   \end{eqnarray}
   We have also plotted the 3-dimensional figure for the payoff of player A in Fig.3.
   We can not find any saddle point. In other word, the $Nash$ $ Equilibrium$ strategy
   disappears in this quantum game. Likewise, when we let $a_{0}=a_{1}=\frac{1}{\sqrt{2}}$,
   $a_{2}=0$, the saddle does not exist either.
   All these show that the \emph{Nash Equilibrium} strategy don't always exist in quantum game
   in an entangled initial state.

   In a general case when player A has $N$ strategies and player B has $M$ strategies,
   we assume the payoff function of player A is
   \begin{equation}
   A=\left[ \begin{array}{ccccc} & \left[ 1\right]  & \left[ 2\right]
    & \ldots & \left[ M\right]  \\ \left[ 1\right]  & \alpha _{11} & \alpha _{12} & \ldots
    & \alpha _{1M} \\ \left[ 2\right]  & \alpha _{21} & \alpha _{11} & \ldots & \alpha _{2M}
     \\ \ldots & \ldots & \ldots & \ldots & \ldots \\ \left[ N\right]  & \alpha _{N1} &
     \alpha _{N2} & \ldots & \alpha _{NM} \end{array} \right]. \label{e14}
     \end{equation}
     Generally, we use the initial state $|\psi _{in}\rangle =\sum\limits_{ijk}a_{k}|ij
     \rangle $, where $\sum\limits_{k}\left\vert a_{k}\right\vert ^{2}=1$; $a_{k}\ge 0$.
     We define the $M$ matrices of player $A$ and player $B$.
     \begin{equation}
     M_{A(N\times N)}=\frac{1}{\sqrt{N-1}}\left[ \begin{array}{cccc} 0 & 1 & \ldots
     & 1 \\ 1 & 0 & \ldots & 1 \\ \ldots & 1 & \ldots & \ldots \\ 1 & 1 & \ldots & 0
     \end{array} \right] ,M_{B(M\times M)}=\frac{1}{\sqrt{M-1}}\left[ \begin{array}{cccc}
     0 & 1 & \ldots & 1 \\ 1 & 0 & \ldots & 1 \\ \ldots & 1 & \ldots & \ldots \\ 1 & 1 &
      \ldots & 0 \end{array} \right]. \label{e15}
      \end{equation}  We see, if we write
      $|1\rangle =\left( 1,0,\ldots0\right) ^T$, $ |2\rangle = \left( 0,1,
      \ldots0\right) ^T$,\ldots $|n\rangle =\left( 0,0,\ldots1\right) ^T$.
       Then we have $M_{A}|1\rangle \longrightarrow |2\rangle +|3\rangle +\ldots+ |n\rangle $,
       \ldots $M_{A}|n\rangle \longrightarrow |1\rangle +|2\rangle +\ldots+|n-1\rangle$.
       It will transform a basis state into the superposition of the rest of the basis states.
       We assume that in the game player $A$ take the unitary operator
       $\sqrt{p}I-\sqrt{1-p}\exp (i\theta _{A})M_{A}$, $\theta _{A}=\arccos (\frac{n-2}
       {2}\sqrt{\frac{ 1-p}{(n-1)p}})$ and  player $B$ take the operator $\sqrt{q}I-
       \sqrt{1-q} \exp (i\theta _{B})M_{B}$, $\theta _{B}=\arccos (\frac{m-2}{2}\sqrt{
       \frac{1-p}{(m-1)p}})$. They can choose a specific value for parameter $p$ and
       $q$. After the operation the state vector becomes
       \begin{equation}
       |\psi _{out}\rangle =C_{11}|11\rangle +C_{12}|12
         \rangle +\ldots+C_{1m}|1m\rangle + C_{21}|21\rangle +\ldots+C_{nm}|nm\rangle,
         \label{e16}
         \end{equation}
 where $\sum\limits_{i,j}^{n,m}\left\vert C_{ij}\right
         \vert ^{2}=1$, and we let  $x_{ij}=\left\vert C_{ij}\right\vert ^{2}$. We
         write the probability matrix
         \begin{equation} X= \left[\begin{array}{cccc} x_{11}
          & x_{12} & \ldots & x_{1m} \\ x_{21} & x_{22} & \ldots & x_{2m} \\ \ldots & \ldots
          & \ldots & \ldots \\ x_{n1} & x_{n2} & \ldots & x_{nm} \end{array}\right].\label{e17}
          \end{equation}
          From equations (\ref{e15},\ref{e16}), we obtain
          \begin{equation} P_{A}(p,q)=
          Tr(XA^T)=\sum\limits_{i,j}^{N,M}x_{ij}a_{ij}. \label{e18}
          \end{equation}
          Now, we let
          \begin{eqnarray} \frac{\partial P_{A}}{\partial p} &=&0, \label{e19} \\
           \frac{\partial P_{A}}{\partial q} &=&0. \nonumber
           \end{eqnarray}
           In term of the
           equation (\ref{e19}), we may obtain the saddle point $p_{0}$, $q_{0}$
         and   the payoff functions
           $P_{A}=P_{A}(p_{0},q_{0})$ and $P_{B}=-P_{A}(p_{0},q_{0})$ at the \emph{Nash
           Equilibrium} point as well.
           It should be emphasized that \emph{Nash Equilibrium} point
            does not always exist in the quantum game. Moreover, it is possible to have
            more than one saddle point in quantum game.

In this paper, we have studied a special multiple strategy
quantum game. It is explicitly demonstrated that entanglement
plays an important role in this quantum game, and it makes the
quantum game different from that of classical game. In
particular, it is found that when the initial state is entangled,
Nash equilibrium does not always exist, in contrast to classical
game. However, when the initial state is a product state, Nash
equilibrium exists. This difference between quantum game and
classical game is because we have set constraint on the allowed
strategies of the two players. When the players are given then
freedom to choose freely the pure strategies, Nash equilibrium
always exists\cite{lee,suncp}. However, for practical purpose, it
is appealing to have a reasonable quantum game with a limited
strategy space. As quantum game theory maybe applicable to many
problems, a restricted strategy quantum game maybe occur, then
one must be careful to examine the strategy set, especially when
the initial state is entangled.

Finally, we thank Mr. Y. S. Li and Miss Liu Fang for help. This work is
    supported in part by China National Science Foundation, the Fok Ying Tung education
    foundation, the National Fundamental Research Program, Contract No. 001CB309308 and
    the Hang-Tian Science foundation.

\begin{figure}
\begin{center}
\includegraphics[width=5cm,height=7cm,angle=-90]{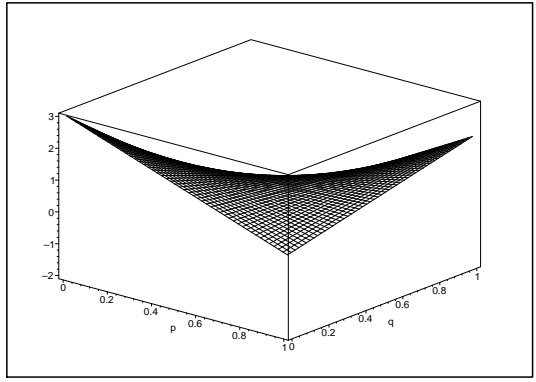}
\caption{The payoff of player A  versus strategy parameters \protect$p\protect$ and
\protect$q\protect$ with initial state \protect$|11\ket\protect$ in a
\protect$2\times 3\protect$ matrix strategy game.}
\end{center}
\end{figure}

\begin{figure}
\begin{center}
\includegraphics[width=5cm,height=7cm,angle=-90]{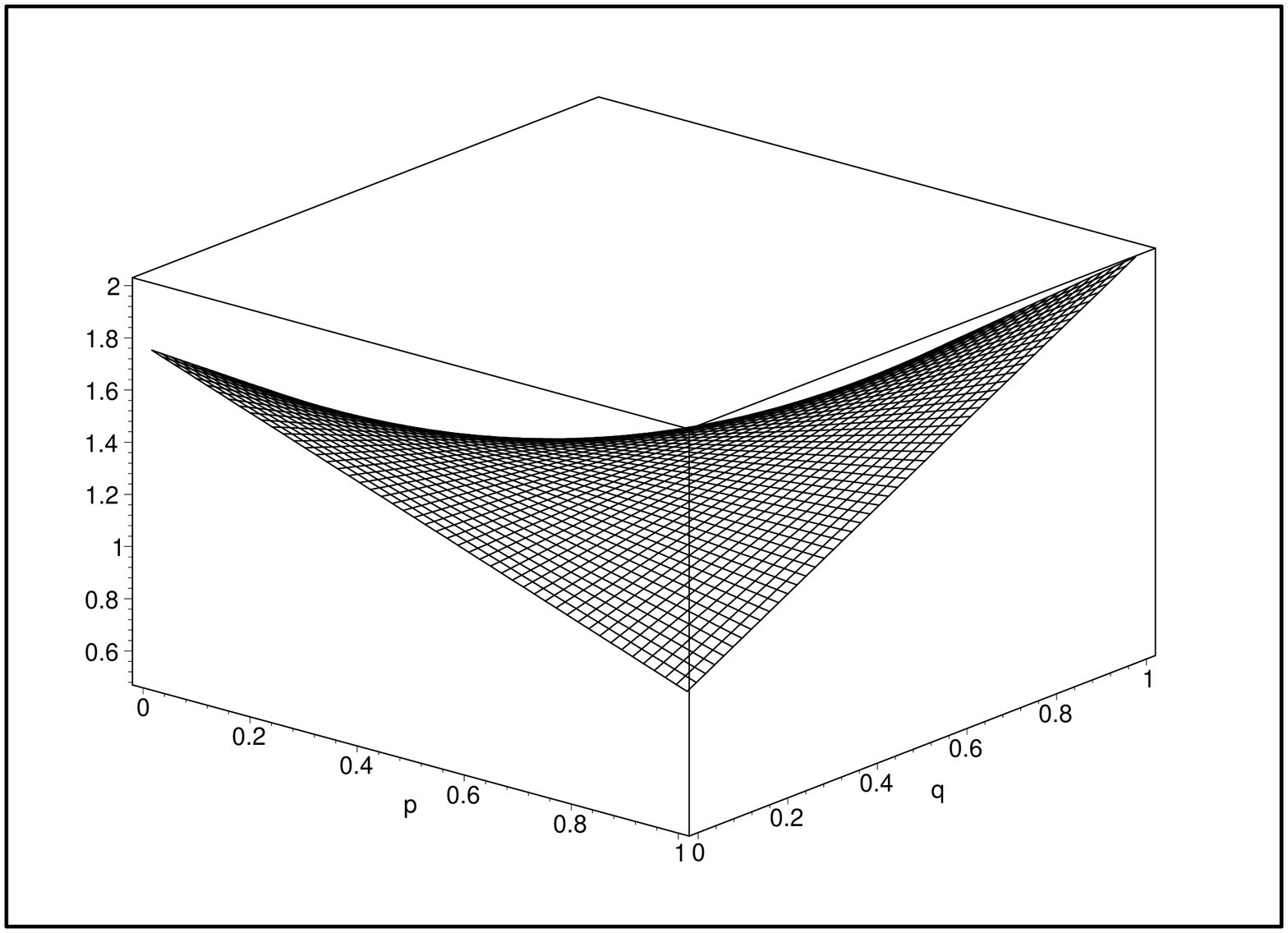}
\caption{The payoff of player A versus strategy parameters \protect$p\protect$ and
\protect$q\protect$ with initial state \protect$|11\ket\protect$ in a \protect$3\times
3\protect$ matrix strategy game. }
\end{center}
\end{figure}

\begin{figure}
\begin{center}
\includegraphics[width=5cm,height=7cm,angle=-90]{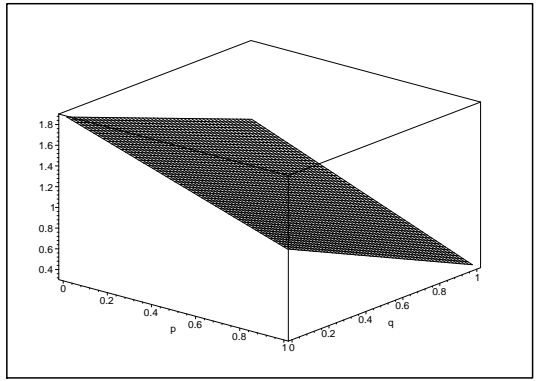}
\caption{The payoff of player A  versus strategy parameters \protect$p \protect$ and
\protect$ q \protect$ with an entangled  initial state \protect$\sqrt{1\over
3}(|11\ket+|22\ket+|33\ket)\protect$ in a \protect$3\times 3\protect$ matrix strategy
game. }
\end{center}
\end{figure}

\begin{thebibliography}{99}
     \bibitem{r1}   J. Eisert,
     M. Wilkens and M. Lewenstein, {\it Phys. Rev. Lett.}{\bf  83},3077(1999)
     \bibitem{r2}   J.
     Eisert, M. Wilkens, {\it J. Mod. Opt.} {\bf  47}, 2543(2000)
      \bibitem{r3}   S. C. Benjamin,
     P. M. Hayden, {\it Phys. Rev.} {\bf A64},030301(2001)
     \bibitem{r4}   A. Iqbal, A.H. Toor, {\it Phys.
     Lett.} {\bf A293},103 (2002)
     \bibitem{r5}   Y. J. Ma, G. L. Long, F. G. Deng, F. Li and S. X. Zhang, {\it Phys. Lett.}
     {\bf A, 301},117 (2002).
     \bibitem{r6}   J. F. Du, et al. {\it Phys. Rev. Lett.} {\bf 88}, 137902 (2002)
     \bibitem{r7}   L. Marinatto and T. Weber, {\it Phys. Lett.} {\bf A272}, 291(2000)
     \bibitem{r8}   A. Iqbal and A. H. Toor, {\it Phys. Lett.} {\bf A280}, 249 (2001)
     \bibitem{r9}  J. von Neumann, O. Morgenstern, Theory of Games and Economic Behaviour,
      Princeton University Press; 3rd edition, 1953.
      \bibitem{lee} C. F. Lee and N. F. Johnson, Lanl-eprint
      quant-ph/0207012
      \bibitem{suncp} X. F. Liu and C. P. Sun, Lanl-eprint,
      quant-ph/0212045
      \end{thebibliography}
      \end{document}